\documentclass[10pt,a4paper]{article}
\usepackage{ioa}
\usepackage{amsmath, amssymb, graphicx, epstopdf,booktabs,array}
\usepackage{subcaption} 
\usepackage{float}
\usepackage[super,sort&compress,comma]{natbib}
\usepackage[utf8]{inputenc}

\usepackage[labelfont=bf,font={small,it}]{caption}
\usepackage{graphicx}
\usepackage[space]{grffile} %allow space character in file names

\graphicspath{{figures/}}

\newcommand{\fixspace}{{\parfillskip=0pt\par}\vspace{-2ex}}
\makeatletter
\renewcommand{\@biblabel}[1]{#1.\hspace{15pt}\hfill}
\makeatother

\usepackage{color}

\begin{document}

% Header formatting stuff
\def\leftmark{}\def\rightmark{}
\def\leftmark{\small \hspace{-2.5mm} {\bf Proceedings of the Institute of Acoustics}}

\title{Comparison of model selection techniques for seafloor scattering statistics
}
\author{}
\vspace{-1.5cm}
\affiliation{
\begin{tabbing}
Derek~R.~Olson\qquad \qquad\  \=Naval Postgraduate School (NPS), Monterey, California, USA \\
Marc~Geilhufe\>   Norwegian Defence Research Establishment (FFI), Kjeller, Norway \\ 
%Email: \> derek.olson@nps.edu
\end{tabbing}
}

%%%%%%%%%%%%%%%%%%%%%%%%%%%%%%%%%%%%%%%%%%%%%%%%%%%%%%%%%%%%%%%%%%%%%
%\abstract
%The Word template does not look like an abstract is desired.

\vspace{1cm}

%%%%%%%%%%%%%%%%%%%%%%%%%%%%%%%%%%%%%%%%%%%%%%%%%%%%%%%%%%%%%%%%%%%%%
\begin{abstract}
	In quantitative analysis of seafloor imagery, it is common to model the collection of individual pixel intensities scattered by the seafloor as a random variable with a given statistical distribution. There is a considerable literature on statistical models for seafloor scattering, mostly focused on areas with statistically homogeneous properties (i.e. exhibiting spatial stationarity). For more complex seafloors, the pixel intensity distribution is more appropriately modeled using a mixture of simple distributions. For very complex seafloors, fitting 3 or more mixture components makes physical sense, but the statistical model becomes much more complex in these cases. Therefore, picking the number of components of the mixture model is a decision that must be made, using a priori information, or using a data driven approach. However, this information is time consuming to collect, and depends on the skill and experience of the human. Therefore, a data-driven approach is advantageous to use, and is explored in this work. Criteria for choosing a model always need to balance the trade-off for the best fit for the data on the one hand and the model complexity on the other hand. In this work, we compare several statistical model selection criteria, e.g., the Bayesian information criterion. Examples are given for SAS data collected by an autonomous underwater vehicle in a rocky environment off the coast of Bergen, Norway using data from the HISAS-1032 synthetic aperture sonar system.
\end{abstract}
\section{Introduction}
\label{sec:introduction}
%%%%%%%%%%%%%%%%%%%%%%%%%%%%%%%%%%%%%%%%%%%%%%%%%%%%%%%%%%%%%%%%%%%%%
Acoustic measurements of seafloor backscattering are a source of unwanted sound in seafloor object detection \cite{Williams2015,Quinn2012,Galusha2018}, but also provide a rich set of information regarding the seafloor properties and structure \cite{Zare2017,Peeples2022,Blondel2009,Stewart1994}. The intensity in a sonar image (i.e. a spatial map of measured backscattering) is typically characterized by a random process \cite{Stewart1994}. There are a variety of metrics, or features, that can be used to describe this random process, including the autocorrelation function, power spectrum \cite{Weszka1976}, wavelet decomposition \cite{Williams2009a}, gray-level co-occurance matrix \cite{Haralick1973,Blondel2009}, the mean intensity (scattering cross section) \cite{Jackson2007,Olson2016} and in general, the intensity probability density function (pdf) \cite{Lyons1999,Olson2019,Gauss2015}.

It was found that for complex scattering environments (such as rocky seafloors, and man-made structures), a mixture pdf was most appropriate\cite{Abraham2011,Olson2019}, which was justified by the non-stationary character of the acoustic data. Each sample of the data was modeled as being drawn from a finite number of distributions, e.g. either from the seafloor or man-made structure, or from horizontal or vertical facets.

In general, the number of components that make up a non-stationary sonar image is unknown, and must be selected prior to choosing a model and estimating the parameters. The more model parameters are used (i.e. more components, or a more complex statistical model for each component), the better the data will be fit, but the parameters may lose meaning. In this work, we explore the use of several model selection techniques based on Bayesian statistics, primarily the Bayesian information criterion (BIC) and Akaike information criterion (AIC). These techniques penalize more complex models in different ways. We also use the log-likelihood  (LL) to characterize the model-data fit.

This paper is organized as follows. A description of the sonar data used in this work and example images are given in Section~\ref{sec:data}. The background statistical modeling and model selection techniques are given in Section~\ref{sec:background}. Results are presented and discussed in Section~\ref{sec:results}. Conclusions are given in Section~\ref{sec:conclusion}.

\section{Data}
\label{sec:data}
The sonar measurements used in this work are synthetic aperture sonar (SAS) images collected off the coast of Bergen, Norway by the Norwegian Defence Research establishment (FFI). The platform used for these measurements is the HUGIN-HUS autonomous underwater vehicle (AUV), using a HISAS-1032 interferometric SAS. This sonar system has a center frequency of 100 kHz and a bandwidth of 30 kHz. The beamformed data is oversampled on a grid with 2$\times$2 cm resolution. Data that is used to fit mixture models is decimated by a factor of 6 in each dimension to reduce the computational complexity of parameter estimation, while reducing the correlation between samples due to the point spread function of the sonar system.

\begin{figure}
\centering
\includegraphics[width=0.9\textwidth]{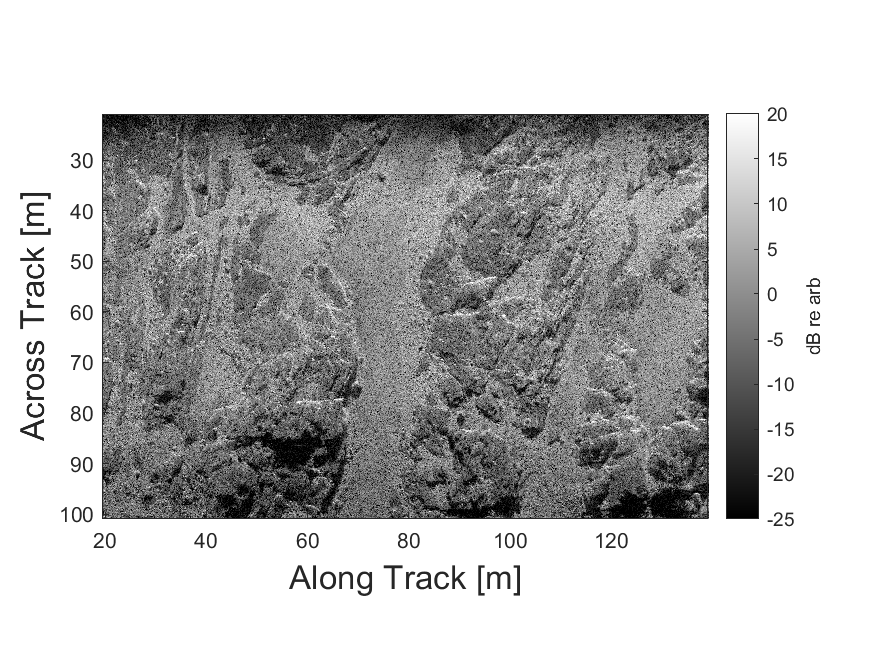}
\fixspace
\fixspace
\fixspace
\fixspace
\caption{An example SAS image, plotted as a function of along track distance on the horizontal axis and across track distance on the vertical axis. The color scale is in decibels referenced to an arbitrary pixel pressure, since the system is uncalibrated.}
\label{fig:full_sas_image}
\end{figure}

\begin{figure}
    \centering
    \includegraphics[width=1.0\textwidth]{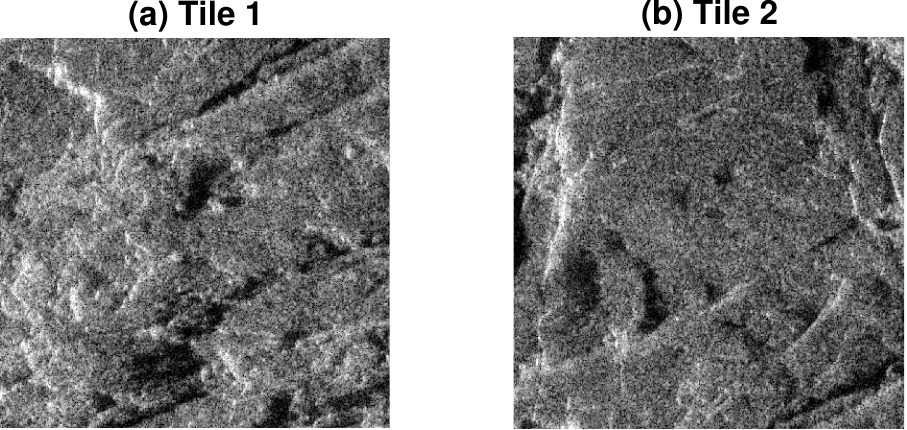}
    \fixspace
    \fixspace
    \caption{Two tiles from Fig.~\ref{fig:full_sas_image}, plotted on a decibel scale with 40 dB of dynamic range. Each image is 600$\times$600 pixels.}
    \label{fig:two-tiles}
\end{figure}

An example image is shown in Fig.~\ref{fig:full_sas_image}. The image consists of an exposed rock outcrop, with sedimented areas in between. To show the detailed environmental structure, two tiles are plotted in Fig.~\ref{fig:two-tiles}, both of which are 600$\times$600 pixels, or 12 m per side.

These tiles show that the rock structure consists of a low intensity uniform scatterer that varies continuously due to undulations in the rock structure. These continuous variations in intensity are punctuated by bright and dark lines due to fractures, and steps created by glacial erosion. These features are distinguishable due to their different intensities, and the SAS system likely has a high enough resolution that discrete scatterers in the environment are distinguishable. Therefore a mixture model is appropriate for modeling the pdf of the ensemble consisting of the pixels from each tile. 

\newpage
\section{Background}
\label{sec:background}
In this section, the basic definitions, statistical models, and model selection definitions are given. We assume data is given in the form of an $N_1\times N_2$ array of intensity values, where the intensity samples are statistically uncorrelated with each other. The total number of samples is $N=N_1 N_2$. The random variable for intensity is denoted $I$, and the amplitude random variable is $A=\sqrt{I}$. We denote the probability of this variable by $p(a)$, with $a$ being a member of the population.

We use two different distributions to build up a mixture model: a Rayleigh distribution for amplitude, and a K-distribution.
% for full paper need some explanation about why we use these models.
% for full paper change definitions to intensity - the math is simpler at least.
A Rayleigh model for amplitude pdf has the form
\begin{align}
    p_R(a|\lambda_0) = 2a/(\lambda_0)e^{-a^2/\lambda_0}, a\geq0,
\end{align}
where $\lambda_0$ is the mean square value of the pdf. The expected value of the intensity is $E[a^2] = \lambda_0$.

A K-distribution for the scattered field amplitude has the form
\begin{align}
    p_K(a|\lambda,\alpha) = 4/(\sqrt{\lambda} \Gamma(\alpha))(a/\sqrt{\lambda})^\alpha K_{\alpha - 1}(2a/\sqrt{\lambda}), a > 0,
\end{align}
where $\Gamma(\cdot)$ is the gamma function, $\lambda$ is the scale parameter of the K distribution, and $\alpha$ is the shape parameter. The expected value of the intensity for this model is $E[I] = \alpha\lambda = \sigma_K$, where $\sigma_K$ is the average intensity. In the results below, the K distribution is parameterized using the pair $(\sigma_K,\alpha)$ rather than the shape parameter. When used in a mixture distribution, the parameters have subscripts to denote which K component the parameters correspond to.

Mixture models are formed by a weighted sum of individual pdf components. The physical meaning of this type of model is that every measurement in a population, or sample, can be identified with one of the $M$ components. The weights of the distributions, $w_m$ are normalized such that $\sum w_m = 1$, and therefore the weights can be interpreted as the fraction of pixels corresponding to each component.

The form of the mixture models used here is
\begin{align}
    p(a|\theta)= w_0 p_R(a|\lambda_0) + \sum\limits_{m=1}^{M-1} p_K(a|\sigma_{Km},\alpha_m)
\end{align}
where $M$ is the number of mixture components, $\theta$ is a vector of length $k$, consisting of the parameters of the model consisting of $w_m$, $\lambda$ and the average powers and shape parameters of the K distributions. The parameters of this mixture model are found using the expectation-maximization (EM) algorithm \cite{Dempster1977}. This method maximizes a slightly altered version of the log-likelihood for each component, but asymptotically maximizes the likelihood function for mixture models \cite{Dempster1977}.

Since the number of components that constitutes the environment is in general unknown, model selection techniques are used to pick the number of K distribution components. As the number of components, $M$, increases, it is better able to match the pdf of the measurement, but at the cost of more uncertainty per parameter \cite{Olson2019}. The likelihood function, $\ell(\theta|a)$, is a common metric for model-data fit. It is defined for N independent samples by, 
\begin{align}
    \ell(\theta|a) = \prod\limits_{n=1}^{N} p(a_n|\theta)
\end{align}
where $a_n$ is the $n-th$ member of the ensemble, and again $\theta$ is the parameter vector. The parameter \fixspace vector $\hat{\theta}$ that maximizes $\ell$ is called the maximum likelihood estimate. It is common to work with the log-likelihood, $\mathcal{L} = \log(\ell)$, which is given as
\begin{align}
    \mathcal{L}(\theta|a) = \sum\limits_{n=1}^{N} \log(p(a_n|\theta))
\end{align}

As stated before, more complex models typically result in a more uncertain maximum likelihood function, so $\mathcal{L}$ cannot be used as a basis on which a decision about $M$ can be made. There exist various model selection techniques, some of which are based on Bayesian concepts\cite{Gelman2013}. Two simple metrics are based on the log-likelihood function, but with an additional penalty that depends on the number of parameters, $k$. The Bayesian information criterion (BIC) is defined as 
\begin{align}
    BIC = -2\mathcal{L}(\hat{\theta}) + \log(N)k
\end{align}
with smaller $BIC$ preferred.
This criterion results from an asymptotic (i.e. large N) Gaussian approximation of the posterior probability density (ppd) function of $\theta$ given the data. The penalty to the log-likelihood is therefore a function of the number of parameters, which is due to the ppd becoming narrower as $N$ becomes asymptotically large. Another model selection criterion is the Akaike information criterion (AIC), due to Akaike \cite{Akaike1974}, and also described by Gelman \cite{Gelman2013}. It is given by the formula
\begin{align}
    AIC = -2\mathcal{L}(\hat{\theta}) + 2k
\end{align}
This criterion is simpler than the BIC and penalizes the log-likelihood independently of the number of samples.

Both the AIC and BIC are based on the log-likelihood function evaluated at $\hat{\theta}$, and are therefore quite sensitive to the numerical estimate. Additionally, point estimates do not contain any information about the ppd as a whole. Other information criteria such as the deviance information criterion and the Wantanabe-Akaike information criterion\cite{Gelman2013} are not studied here, but we consider them as fruitful areas for future work. Another possibility is to use trans-dimensional Monte-Carlo methods to estimate the model with the highest posterior probability \cite{Dettmer2010}.

%The last criterion studied here is the deviance information criterion, which is based on a Monte-carlo sampling of the 

\section{Results and Discussion}
\label{sec:results}
\begin{figure}
     \centering
     \includegraphics[width=1.0\textwidth]{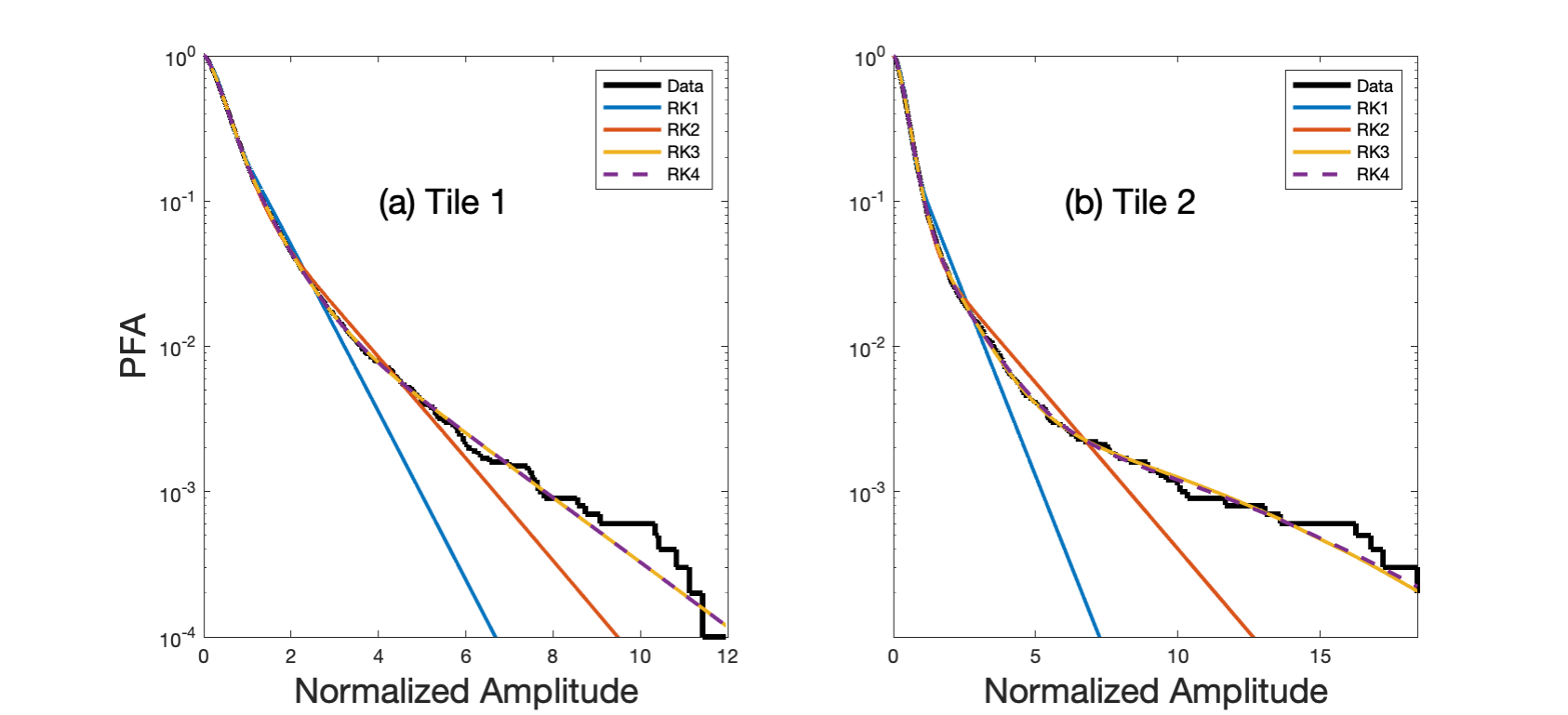}
     \fixspace
     \fixspace
     \caption{The probability of false alarm (PFA) for the data from snippets 1 and 2, compared to the various mixture models explored here.}
     \label{fig:model_pfas}
 \end{figure}

R-K mixture models were fit to the amplitude data in Figs.~\ref{fig:two-tiles} (a) and (b), using between $M=2$ and $M=5$ components. This means that the maximum number of K-distribution components was 4. The model-data fit is shown graphically in Fig.~\ref{fig:model_pfas} for both tiles in terms of the log of the probability of false alarm (PFA), also called the excedence distribution function (EDF). The PFA is related to the cumulative distribution function (CDF) through $PFA=1-CDF$, and is a common method of presenting sonar reverberation statistics \cite{Lyons2009}. The data in Tile 1 shows a slight ``knee'' near a normalized amplitude of 2, 6, and 8. These changes in slope of the log-PFA indicate different components that make up the model. The log-PFA of Tile 2 has more pronounced knees in the curve, near the normalized amplitudes of 2, 4, and 12, although the last one is more uncertain due to the finite sample size being more evident at high amplitudes (i.e. the PFA curve becomes more stair-case like, rather than a smooth curve).

Model-data fits for both tiles are poor for both R-K1 and R-K2. This behavior is likely due to the fact that there are not enough components to fit the data. R-K3 and R-K4 fit the data much better, but are almost the same. This behavior indicates that a 5-component model does not provide significantly better fit than the 4-component model. From a visual standpoint, it makes sense to use a 4-component model for both of these data sets. This hypothesis will be compared to the results of the more formal model selection techniques.

Model section criteria, the log-likelihood (LL), BIC, and AIC are shown for both tiles in Table~\ref{tab:model_selection_results}. The maximum LL value for Tile 1 is a tie between the 4- and 5-component models, and the maximum LL for Tile 2 is for the 4-component model. Intuitively, the more complex model should have a higher likelihood function, but here, this may be an issue with the numerical parameter estimates. Additionally, as discussed in Section~\ref{sec:background}, one cannot base a model selection decision purely on the maximum of the likelihood function. Further work on refining these estimate should be made. In terms of the AIC, the smallest value occurs for the 4-component model for both datasets. However, for the BIC, the smallest value occurs for the 3-component model for Tile 2, and the 4-component model for Tile 1. The AIC penalizes model complexity only slightly, whereas the BIC has a severe penalty for complexity for large number of data samples. Since the number of data samples was about $10^4$, the penalty for model complexity is about $0.5 \log10^4\approx 4.6$ times larger for the BIC as it is for the AIC. We conclude that the 3-component model is preferred by BIC for Tile 2 because of the much larger penalty for model complexity.

\begin{table}[]
    \centering
    \begin{tabular}{c|c c c | c c c }
    \hline \hline
             & \multicolumn{3}{c|}{Tile 1} & \multicolumn{3}{c}{Tile 2} \\
      Model  & LL   & AIC & BIC & LL & AIC & BIC  \\
      \hline
       R-K1  & -4321 &  8652  & 8688 & -885 & 1780 & 1810 \\
       R-K2  & -4157 &  8331  & 8389 & -847 & 1711 & 1759   \\
       R-K3  & -4137 &  8297  & 8377 & -839 & 1700 & 1766  \\
       R-K4  & -4137 &  8301  & 8404 & -840 & 1709 & 1793  \\
       \hline \hline
    \end{tabular}
    \fixspace
    \caption{Model selection results for both image tiles. The log-likelihood (LL), Akaike information criterion (AIC), and the Bayesian information criterion (BIC) are shown.}
    \label{tab:model_selection_results}
\end{table}

\section{Conclusion}
\label{sec:conclusion}
We presented a statistical model for SAS images of complex, non-stationary, rocky seafloors. This model consisted of a Rayleigh distribution, plus an unknown number of K distributions. The number of K-distributions was selected using model selection techniques, including the AIC and BIC. For the AIC, the 4-component model was selected as the most appropriate for two image tiles used here. For the BIC, different number of components were chosen for each image tile. It is likely that a different number of components was chosen by the BIC due to its more severe penalty for model complexity. Future work should include more robust model selections techniques, such as the deviance information criterion, which employs a Monte-Carlo Markov chain sampling of the distribution parameters. This work can also be used to partition an image into different scattering components, which can aid in estimates of image complexity, and may also be the basis for quantitative seafloor remote sensing of geological and/or biological parameters.

\section{Acknowledgments}
This research was supported by grant number N00014-23-WX01149 from the Office of Naval Research under the Young Investigator Program. The authors thank FFI's HUGIN-HUS operator group for collecting the data used in this study.
%%%%%%%%%%%%%%%%%%%%%%%%%%%%%%%%%%%%%%%%%%%%%%%%%%%%%%%%%%%%%%%%%%%%%
%\bibliographystyle{IEEEtran}
\renewcommand\refname{REFERENCES\vspace{10pt}}
\renewcommand\bibsection{\section{\refname}}

%\bibliography{derek_bibilography}

\begin{thebibliography}{10}
\providecommand{\url}[1]{#1}
\csname url@samestyle\endcsname
\providecommand{\newblock}{\relax}
\providecommand{\bibinfo}[2]{#2}
\providecommand{\BIBentrySTDinterwordspacing}{\spaceskip=0pt\relax}
\providecommand{\BIBentryALTinterwordstretchfactor}{4}
\providecommand{\BIBentryALTinterwordspacing}{\spaceskip=\fontdimen2\font plus
\BIBentryALTinterwordstretchfactor\fontdimen3\font minus
  \fontdimen4\font\relax}
\providecommand{\BIBforeignlanguage}[2]{{%
\expandafter\ifx\csname l@#1\endcsname\relax
\typeout{** WARNING: IEEEtran.bst: No hyphenation pattern has been}%
\typeout{** loaded for the language `#1'. Using the pattern for}%
\typeout{** the default language instead.}%
\else
\language=\csname l@#1\endcsname
\fi
#2}}
\providecommand{\BIBdecl}{\relax}
\BIBdecl

\bibitem{Williams2015}
D.~P. Williams, ``Fast unsupervised seafloor characterization in sonar imagery
  using lacunarity,'' \emph{IEEE Transactions on Geoscience and Remote
  Sensing}, vol.~53, no.~11, pp. 6022--6034, Nov 2015.

\bibitem{Quinn2012}
R.~Quinn, \emph{Acoustic Remote Sensing in Maritime Archaeology}, B.~Ford,
  D.~L. Hamilton, and A.~Catsambis, Eds.\hskip 1em plus 0.5em minus 0.4em\relax
  Oxford University Press, sep 2012.

\bibitem{Galusha2018}
A.~P. Galusha, J.~M. Keller, A.~Zare, and G.~Galusha, ``A fast target detection
  algorithm for underwater synthetic aperture sonar imagery,'' in
  \emph{Detection and Sensing of Mines, Explosive Objects, and Obscured Targets
  {XXIII}}, J.~C. Isaacs and S.~S. Bishop, Eds.\hskip 1em plus 0.5em minus
  0.4em\relax {SPIE}, apr 2018.

\bibitem{Zare2017}
A.~Zare, N.~Young, D.~Suen, T.~Nabelek, A.~Galusha, and J.~Keller,
  ``Possibilistic fuzzy local information c-means for sonar image
  segmentation,'' in \emph{2017 {IEEE} Symposium Series on Computational
  Intelligence ({SSCI})}.\hskip 1em plus 0.5em minus 0.4em\relax {IEEE}, nov
  2017.

\bibitem{Peeples2022}
J.~Peeples, W.~Xu, and A.~Zare, ``Histogram layers for texture analysis,''
  \emph{{IEEE} Transactions on Artificial Intelligence}, vol.~3, no.~4, pp.
  541--552, aug 2022.

\bibitem{Blondel2009}
P.~Blondel and O.~G. Sichi, ``Textural analyses of multibeam sonar imagery from
  {Stanton Banks}, {Northern Ireland} continental shelf,'' \emph{Applied
  Acoustics}, vol.~70, no.~10, pp. 1288--1297, oct 2009.

\bibitem{Stewart1994}
W.~K. Stewart, D.~Chu, A.~Malik, S.~Lerner, and H.~Singh, ``Quantitative
  seafloor characterization using a bathymetric sidescan sonar,'' \emph{IEEE J.
  Ocean Eng.}, vol.~19, pp. 599--610, 1994.

\bibitem{Weszka1976}
J.~S. Weszka, C.~R. Dyer, and A.~Rosenfeld, ``A comparative study of texture
  measures for terrain classification,'' \emph{{IEEE} Transactions on Systems,
  Man, and Cybernetics}, vol. {SMC}-6, no.~4, pp. 269--285, 1976.

\bibitem{Williams2009a}
D.~P. Williams, ``Unsupervised seabed segmentation of synthetic aperture sonar
  imagery via wavelet features and spectral clustering,'' in \emph{2009 16th
  {IEEE} International Conference on Image Processing ({ICIP})}.\hskip 1em plus
  0.5em minus 0.4em\relax {IEEE}, nov 2009.

\bibitem{Haralick1973}
R.~M. Haralick, K.~Shanmugam, and I.~Dinstein, ``Textural features for image
  classification,'' \emph{{IEEE} Transactions on Systems, Man, and
  Cybernetics}, vol. {SMC}-3, no.~6, pp. 610--621, nov 1973.

\bibitem{Jackson2007}
D.~R. Jackson and M.~D. Richardson, \emph{High-Frequency Seafloor
  Acoustics}.\hskip 1em plus 0.5em minus 0.4em\relax New York, NY: Springer,
  2007.

\bibitem{Olson2016}
D.~R. Olson, A.~P. Lyons, and T.~O. S{\ae}b{\o}, ``Measurements of
  high-frequency acoustic scattering from glacially eroded rock outcrops,''
  \emph{J. Acoust. Soc. Am.}, vol. 139, no.~4, pp. 1833--1847, 2016.

\bibitem{Lyons1999}
A.~P. Lyons and D.~A. Abraham, ``Statistical characterization of high-frequency
  shallow-water seafloor backscatter,'' \emph{The Journal of the Acoustical
  Society of America}, vol. 106, no.~3, pp. 1307--1315, sep 1999.

\bibitem{Olson2019}
D.~R. Olson, A.~P. Lyons, D.~A. Abraham, and T.~O. Sæbø, ``Scattering
  statistics of rock outcrops: Model-data comparisons and bayesian inference
  using mixture distributions,'' \emph{J. Acoust. Soc. Am.}, vol. 145, no.~2,
  pp. 761--774, 2019.

\bibitem{Gauss2015}
R.~C. Gauss, J.~M. Fialkowski, D.~C. Calvo, R.~Menis, D.~R. Olson, and A.~P.
  Lyons, ``Moment-based method to statistically categorize rock outcrops based
  on their topographical features,'' in \emph{OCEANS 2015 - MTS/IEEE
  Washington}, Oct 2015, pp. 1--5.

\bibitem{Abraham2011}
D.~Abraham, J.~Gelb, and A.~Oldag, ``Background and clutter mixture
  distributions for active sonar statistics,'' \emph{IEEE J. Ocean Eng.},
  vol.~36, no.~2, pp. 231--247, April 2011.

\bibitem{Dempster1977}
A.~P. Dempster, N.~M. Laird, and D.~B. Rubin, ``Maximum likelihood from
  incomplete data via the {EM} algorithm,'' \emph{Journal of the Royal
  Statistical Society: Series B (Methodological)}, vol.~39, no.~1, pp. 1--22,
  sep 1977.

\bibitem{Gelman2013}
A.~Gelman, J.~B. Carlin, H.~S. Stern, D.~B. Dunson, A.~Vehtari, and D.~B.
  Rubin, \emph{Bayesian Data Analysis}.\hskip 1em plus 0.5em minus 0.4em\relax
  Chapman and Hall/{CRC}, nov 2013.

\bibitem{Akaike1974}
H.~Akaike, ``A new look at the statistical model identification,'' \emph{{IEEE}
  Transactions on Automatic Control}, vol.~19, no.~6, pp. 716--723, dec 1974.

\bibitem{Dettmer2010}
J.~Dettmer, S.~E. Dosso, and C.~W. Holland, ``Trans-dimensional geoacoustic
  inversion,'' \emph{The Journal of the Acoustical Society of America}, vol.
  128, no.~6, pp. 3393--3405, dec 2010.

\bibitem{Lyons2009}
A.~Lyons, S.~Johnson, D.~Abraham, and E.~Pouliquen, ``High-frequency scattered
  envelope statistics of patchy seafloors,'' \emph{{IEEE} Journal of Oceanic
  Engineering}, vol.~34, no.~4, pp. 451--458, oct 2009.

\end{thebibliography}

\end{document}